# The Role of Hypersound in Stimulation of Resonant Transitions in Positronium Atom

E.A. Ayryan [1] , A.S. Gevorkyan [2,3] , M. Hnatic[4,5,6] , K.B. Oganesyan [1,7*] , Yu.V. Rostovtsev [8]

[*] bsk@yerphi.am

[1] LIT, Joint Institute for Nuclear Research, Dubna, Russia
[2] Institute for Informatics and Automation Problems,
[3] Institute for Chemical Physics, NAS of Armenia,Yerevan, Armenia,
[4] BLTP, Joint Institute for Nuclear Research, Dubna, Russia

[5] Faculty of Sciences, P. J. Safarik University, Kosice, Slovakia

[6] Institute of Experimental Physics SAS, Kosice, Slovakia

[7] A.I. Alikhanyan National Science Lab, Yerevan Physics Institute, Yerevan, Armenia,

[8] University of North Texas, Denton, TX, USA,

**Abstract.** The possibilities of stimulation of resonant transitions between quantum states of Positron Atom by the external hypersound are investigated in detail.

## 1. Introduction

The annihilation of positrons in ionic crystals is developing direction (see [1-4] and references therein). This is way to get short wavelength radiation. The idea is based on channeling of positrons in ionic crystals like CsCl. In papers [1-4] a possibility of formation of stable positrinium atoms (PA) in such systems was demonstrated. One can find many aspects also in [5-38].

In the present paper the possibilities of stimulation of resonant transitions between quantum states of Positron Atom by the external hypersound are investigated in detail.

## 2. The Transition Probability Between Different Quantum States

Under the influence of longitudinal hypersonic oscillations in the crystal lattice along the channeling axis z, a frequency-modulated potential is being formed [1]:

$$U(z,\rho) = (1+\sigma\cos(k_s z))U_0(\rho) + \delta\cos(k_s z), \tag{1}$$

where $k_s = 2\pi/\lambda_s$, $\lambda_s$ is the wavelength of the hypersound, and $\sigma$ , $\delta$ are magnitudes describing the modulation of the potential.

For the considered case the positron wavefunction reads:

$$\Psi(r) = \frac{1}{(2\pi\lambda_s)^{1/2}} \exp\left( i\int_0^z p(z')dz' \right)\Phi(z,\rho,\varphi), \tag{2}$$

where $p(z) = \{2\mu[E - U(z,\rho_0)]\}^{1/2}$ and E is a positron total energy. Now substituting representation (2) into the 3D Schrodinger equation in the limit of the quasiclassical approximation, it is easy to find the following parabolic equation for the wavefunction:

$$\begin{aligned} & i\frac{\partial\Phi}{\partial\tau} = \hat{H}(\tau,\rho,\varphi), \quad \tau = \int_0^z \frac{dz'}{p(z')}, \\ & \hat{H} = -\frac{1}{\rho}\frac{\partial}{\partial\rho}\left(\rho\frac{\partial}{\partial\rho}\right) - \frac{1}{\rho^2}\frac{\partial^2}{\partial\varphi^2} - 2\mu[U(z,\rho_0) - U(z,\rho)]. \end{aligned} \tag{3}$$

It is obvious that Hamiltonian (3) satisfies the periodic condition:

$$\hat{H}(\tau + T) = \hat{H}(\tau), \quad T = \int_0^{\lambda_s} p^{-1}(z')dz';$$

correspondingly the solution of Equation (3) may be represented in the form [1,2]:

$$\Phi(\tau,\rho,\varphi) = \exp(-i\varepsilon\tau)\Phi^{(\varepsilon)}(\tau,\rho,\varphi), \quad \Phi^{(\varepsilon)}(\tau + T) = \Phi^{(\varepsilon)}(\tau), \tag{4}$$

where $\varepsilon = \varepsilon^{(0)} + U_0(\rho_0)$ is a quasienergy.

Further, representing the wavefunction as:

$$\begin{aligned} & \Phi^{(\varepsilon)}(\tau,\rho,\varphi) = \frac{1}{(2\pi\rho)^{1/2}}\exp\left(im\varphi\right)X^{(\varepsilon)}(z,\rho), \\ & m = 0, \pm 1, \pm 2, ..., \end{aligned} \tag{5}$$

Subject to (4) we can obtain the following equation:

$$\left\{ i\frac{\partial}{\partial\tau} + \frac{\partial^2}{\partial\tau^2} + 2\mu\left[\varepsilon - \frac{M}{2\mu\rho^2} + U(z,\rho_0) - U(z,\rho)\right]\right\}X^{(\varepsilon)} = 0. \tag{6}$$

The solution of Equation (6) may be represented in the form of expansion:

$$X_{nm}^{(\varepsilon)}(\tau,\rho)=\frac{1}{(2\pi)^{1/2}}\sum_{k=1}^{k_0}C_{knm}(\tau)X_{km}^{(0)}(\rho), \qquad (7)$$

where $k_0$ is a maximal value of the vibration quantum number. Note that in absence of hypersound the coefficients $C_{knm}(\tau)$ are given:

$$C_{knm}^{(0)}(\tau)=\delta_{kn}\lambda_s^{-1/2}\exp(ip_z z),$$

and Equation (6) transforms to Equation (15) of [1], correspondingly the solution (7) transforms to solution (17) of [1].

Substituting expression (7) into Equation (6) and using the orthonormal behavior of wavefunction $X_{km}^{(0)}(\rho)$ (see expressions (14) (17) and (19) of [1]), we can receive the following system of first order ordinary differential equations:

$$i\frac{d\hat{C}_{knm}(z)}{dz}=\frac{\sigma\cos(k_s z)}{(2\pi)^{1/2}p(z)}\sum_{l=1}^{k_0}\Lambda_{kl}\hat{C}_{knm}(z), \qquad (8)$$

where $\hat{C}_{knm}(z)=C_{knm}(\tau)$; in addition:

$$\Lambda_{kl}=-U_0(\rho_0)\delta_{kl}+\int_0^{\infty}U_0(\rho)X_{km}^{(0)}(\rho)X_{lm}^{(0)}(\rho)d\rho.$$

Recall that for solving the system of Equation (8) it is necessary to have in view of the normalization condition:

$$\sum_{l=1}^{k_0}|\hat{C}_{knm}|^2=\lambda_s^{-1}.$$

As was shown in [39] Equation (8) at certain wavelengths of hypersound has a selective solution, i.e. except for one coefficient $\hat{C}_{knm}$, all others are suppressed. In particular it is shown that at condition $\lambda_s=\lambda_{kn}=2\pi|\omega_{kn}|^{-1}$, where $\omega_{kn}=\varepsilon_{km}^{(0)}-\varepsilon_{nm}^{(0)}$ ($k\neq n$, k,n = 1, 2, 3, . . .), the hypersound influence on a positron atom is resonant.

Finally if we neglect the impulse recoil of a photon, which in the considered case is justified, the transition probability between different quantum states on the length $\lambda_s$ will be defined by the formula:

$$P_{kn}^{m}=\frac{1}{\lambda_s}\left|\int_0^{\lambda_s}\hat{C}_{knm}(z')dz'\right|,$$

where m describes the quantum number of the rotation state of PA that doesn't change at quantum transition.

## 3. Conclusion

In conclusion it is necessary to note that along the <100> axis, the positron can be induced into the superchanneling regime. This phenomenon allows us to use hypersound for the stimulation of resonant transitions between quantum states of a positron atom with great effectiveness and correspondingly strengthens the radiation's effect of channeling substantially.

## Acknowledgements

KBO and EAA are grateful for the support from the NSP of Slovak Republic.